\documentclass{Interspeech2024}
\usepackage{hyperref}
\usepackage{tabularray}
\usepackage{adjustbox}
\usepackage{graphicx}
\usepackage{multirow}
\usepackage{hyperref}
\usepackage{cite}

\interspeechcameraready

\title{Vec-Tok-VC+: Residual-enhanced Robust Zero-shot Voice Conversion with Progressive Constraints in a Dual-mode Training Strategy}

\name[affiliation={1}]{Linhan}{Ma}
\name[affiliation={1}]{Xinfa}{Zhu}
\name[affiliation={1}]{Yuanjun}{Lv}
\name[affiliation={1}]{Zhichao}{Wang}
\name[affiliation={1}]{Ziqian}{Wang}
\name[affiliation={2}]{Wendi}{He}
\name[affiliation={2}]{Hongbin}{Zhou}
\name[affiliation={1,*}]{Lei}{Xie}

\address{
  $^1$Audio, Speech and Language Processing Group (ASLP@NPU), \\ School of Computer Science, Northwestern Polytechnical University, Xi'an, China\\
  $^2$Ximalaya Inc, China}
\email{mlh2023@mail.nwpu.edu.cn, lxie@nwpu.edu.cn\thanks{* Corresponding author.}}


\begin{document}

\maketitle

\begin{abstract}
    Zero-shot voice conversion (VC) aims to transform source speech into arbitrary unseen target voice while keeping the linguistic content unchanged. 
    Recent VC methods have made significant progress, but semantic losses in the decoupling process as well as training-inference mismatch still hinder conversion performance.
    In this paper, we propose Vec-Tok-VC+, a novel prompt-based zero-shot VC model improved from Vec-Tok Codec, achieving voice conversion given only a 3s target speaker prompt. 
    We design a residual-enhanced K-Means decoupler to enhance the semantic content extraction with a two-layer clustering process. 
    Besides, we employ teacher-guided refinement to simulate the conversion process to eliminate the training-inference mismatch, forming a dual-mode training strategy. 
    Furthermore, we design a multi-codebook progressive loss function to constrain the layer-wise output of the model from coarse to fine to improve speaker similarity and content accuracy.
    Objective and subjective evaluations demonstrate that Vec-Tok-VC+ outperforms the strong baselines in naturalness, intelligibility, and speaker similarity.

\end{abstract}

\section{Introduction}
\label{sec:intro}

Voice conversion (VC) aims to transfer speech from a source speaker to sound like that of a target speaker while keeping the linguistic content unchanged~\cite{VCoverview}. 
VC has been deployed in many applications, such as privacy protection~\cite{yjx_anoy}, movie dubbing~\cite{expressiveVC}, etc. However, these VC systems are limited to converting between predefined speakers and require a sizable amount of the target speaker's speech. Because of the high cost of data collection, achieving conversion with low data requirements of the target speaker is more practical for real-world deployment.

\textit{Zero-shot} VC focuses on converting the speaker timbre of the source speech to that of arbitrary speakers with only one utterance, which has drawn much attention recently. 
Its main challenge lies in modeling unseen speakers' timbre and decoupling the source semantic content.
The popular framework of zero-shot VC is to decompose source speech into speaker timbre and semantic content, consisting of linguistic information and speaking variation, and then convert the speaker timbre of the source speaker to the target speaker. Many approaches have been proposed for zero-shot voice conversion by employing specific-designed structures~\cite{multilevel_structure_vc, oneshotvc_IN}, loss functions~\cite{vqmivc, lossdesign_vc}, and training strategies~\cite{avqvc, strategy2}. For example, some studies~\cite{vqvc, AutoVC, vaevc} incorporate information bottleneck to separate speaker timbre from content representation. Adversarial training~\cite{cycleGANvc, lossdesign_vc} and mutual information constraint~\cite{vqmivc} are also used to reduce the correlations among different speech factors. However, these disentanglement approaches often suffer from the inevitable trade-off between speech quality and speaker similarity. 

For accurately modeling speaker timbre information, some studies~\cite{feagmentVC,vqvc,attn-based-spkemb,hier-spkemb} capture speaker timbre from the multi-reference speech in a finer-grained way to obtain multi-level or time-varying representations. 
Instead of using explicit disentanglement designs in VC training~\cite{oneshotvc_IN, vqmivc}, another popular way is to achieve this before the training. Some studies~\cite{Perturb1, Perturb2} use signal perturbation techniques to alter the pitch and timbre of speech utterances to make pseudo-parallel pairs. With the speaker identity supervision, the speaker verification (SV)~\cite{xvectors} model is leveraged to extract speaker representation, while the automatic speech recognition (ASR) model is employed to extract the content. 
However, the limited capacity of most previous VC models~\cite{expressiveVC,vqmivc,feagmentVC} makes it difficult to leverage large amounts of data, hindering the previous approaches to achieve high-quality conversion on wild unseen speakers.
Besides, some self-supervised learning (SSL) models, such as HuBERT~\cite{Hubert} and WavLM~\cite{WavLM}, can capture the local general structure from speech utterance to form SSL features, which is used in zero-shot VC~\cite{LMVC, SEF-VC}. 
Since the continuous SSL feature captures phonetic similarity and preserves semantic content and speaker information~\cite{WavLM}, kNN-VC~\cite{KNNvc} introduces k-nearest neighbors (kNN) to directly replace the SSL feature of the source speech with that of the target speaker's speech based on feature similarity and to achieve VC. It can achieve high speaker similarity but requires several minutes of speaker's utterances as a matching set. 

Another recent novel method~\cite{vectok} is to use a designed codec structure called Vec-Tok Codec to achieve zero-shot VC.
The basic idea of Vec-Tok-VC is to first represent speech to continuous acoustic and discrete semantic features by SSL model and K-Means clustering quantification, respectively, and then combine the source semantic content and the speaker timbre from the acoustic feature of the target speaker to generate the converted speech.
Benefiting from the scaling up of training data and powerful modeling ability, Vec-Tok Codec can capture speaker timbre from the acoustic feature prompt.
However, its decoupling process based on 300-category K-Means clustering may lose the speaking variations and hurt the linguistic content of the source speech, leading to poor naturalness and content accuracy.
Moreover, in most VC methods including Vec-Tok-VC, different from the inference, the training process uses the target speaker reference and source semantic both from the same utterance. 
This mismatching behavior makes it hard to ensure the decoupling of speaker timbre and content information during training, causing performance degradation~\cite{stylevc, DDDMvc}.

To address these issues, we propose Vec-Tok-VC+, a prompt-based robust zero-shot VC model improved from Vec-Tok Codec integrating a residual-enhanced K-Means decoupler, which converts the enhanced semantic features to target speech condition on 3s target speaker prompt. 
Specifically, inspired by residual vector quantization (RVQ)~\cite{soundstream, encodec}, we incorporate residual-enhanced K-Means quantization to encode the residual information of linguistic content and rich speaking variation to enhance the semantic content,  alleviate the loss of semantic content during the decoupling and enhance the para-linguistic information.  
To obtain better decoupling ability and eliminate the training-inference mismatch, we introduce a teacher-guided refinement process to form a dual-mode (conversion mode and reconstruction mode) training strategy with the original reconstruction process. 
Furthermore, a multi-codebook loss is introduced to help the model fit into the target speech progressively from coarse-grained to fine-grained, to prevent the information dispersed during the multi-layer modeling. Experimental results demonstrate that Vec-Tok-VC+ achieves superior performance over previous zero-shot models in both speaker similarity and speech naturalness. The samples of our proposed systems can be found in our demo page~\footnote{\hyperlink{https://ma-linhan.github.io/VecTokVC-Plus/}{https://ma-linhan.github.io/VecTokVC-Plus/}}.

\section{Proposed approach}

\subsection{System overview}
The Vec-Tok Codec~\cite{vectok} uses WavLM to extract continuous acoustic features and decouples semantic features from the acoustic features through a 300-category K-Means clustering. 
Based on this, Vec-Tok-VC concatenates the acoustic feature prompt of the target speaker and the semantic feature of the source speech along the temporal axe and fed into the conformer-based converter to achieve zero-shot voice conversion.

\begin{figure}[h]
    \centering
    \includegraphics[width=\linewidth]{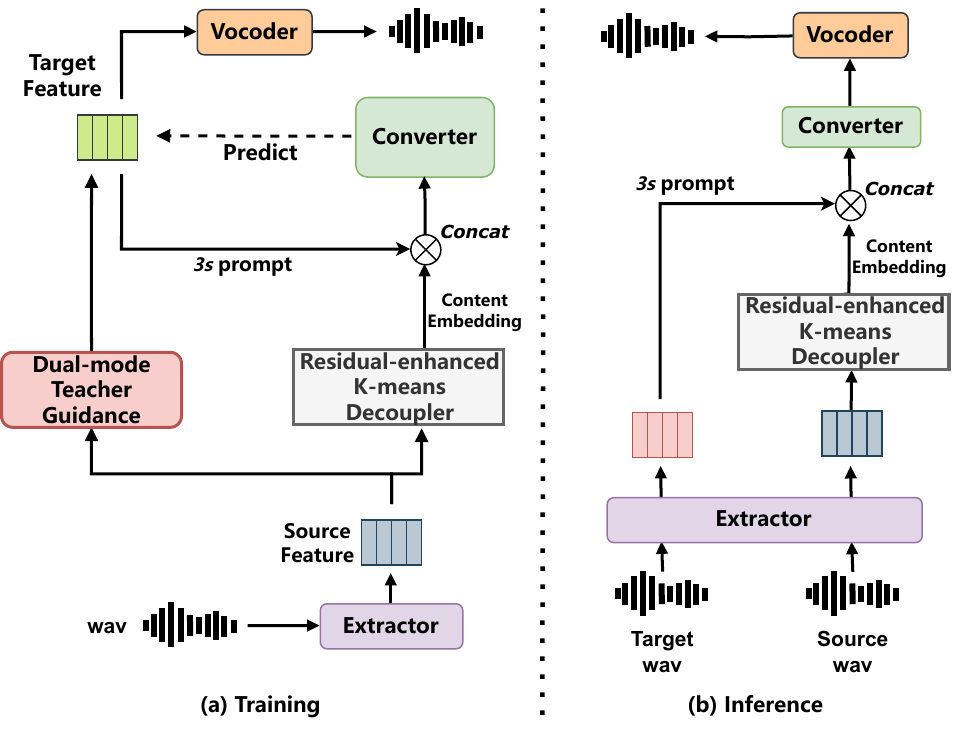}
    \caption{The overview of Vec-Tok-VC+.}
    \label{overview}
\end{figure}

As shown in Fig.~\ref{overview}, improved from Vec-Tok-VC, Vec-Tok-VC+ mainly consists of three parts: a residual-enhanced K-Means decoupler, a prompt-based conformer converter, and a teacher module. At first, the feature extractor extracts the continuous SSL feature, which contains semantic content and speaker timbre information. The Vec-Tok-VC+ is built based on this SSL feature. To squeeze out the speaker timbre from the content information, the decoupler incorporates a residual-enhanced design to K-Means quantization to get enhanced content representations. Taking a short clip of a speaker utterance as a speaker prompt and content representation from the source speech, the conformer converter predicts the target SSL feature. Besides, to mitigate the training inference mismatch, a teacher module is introduced in our framework during training. Finally, a modified HIFIGAN vocoder~\cite{HiFIGAN} is adapted to reconstruct waveform from SSL features.

\textbf{Feature extraction:} In our system, instead of using spectrogram or speech codec, continuous SSL features from the XLSR model, which is a powerful multi-lingual variant of wav2vec 2.0~\cite{w2v2.0}, are selected to represent speech. Previous work~\cite{KNNvc, vectok} has demonstrated that SSL features, 
such as WavLM~\cite{WavLM} and wav2vec 2.0~\cite{w2v2.0},
can be directly used to achieve high-quality speech reconstruction because of its richness of semantic and speaker information. Consequently, the XLSR-base vocoder is introduced to reconstruct the waveform from SSL features.

\textbf{Training stage:} As shown in Fig.~\ref{overview}(a), Vec-Tok-VC+ is trained to convert the source SSL feature to the target SSL feature condition on the explicitly given speaker prompt. During training, the source and target SSL features 
have the same content,
while the target SSL feature is generated by the teacher module when the teacher guidance is activated (See Section~\ref{sec:teacher}). The speaker prompt is randomly selected from the sequence of target features.

\textbf{Zero-shot inference:} in Fig.~\ref{overview}(b), given the SSL feature from the utterance of the target speaker as speaker prompt, Vec-Tok-VC+ outputs the converted speech with source semantic content and target speaker timbre.

\subsection{Residual-enhanced K-Means decoupler}

\begin{figure*}[h]
    \centering
    \includegraphics[width=0.9\linewidth]{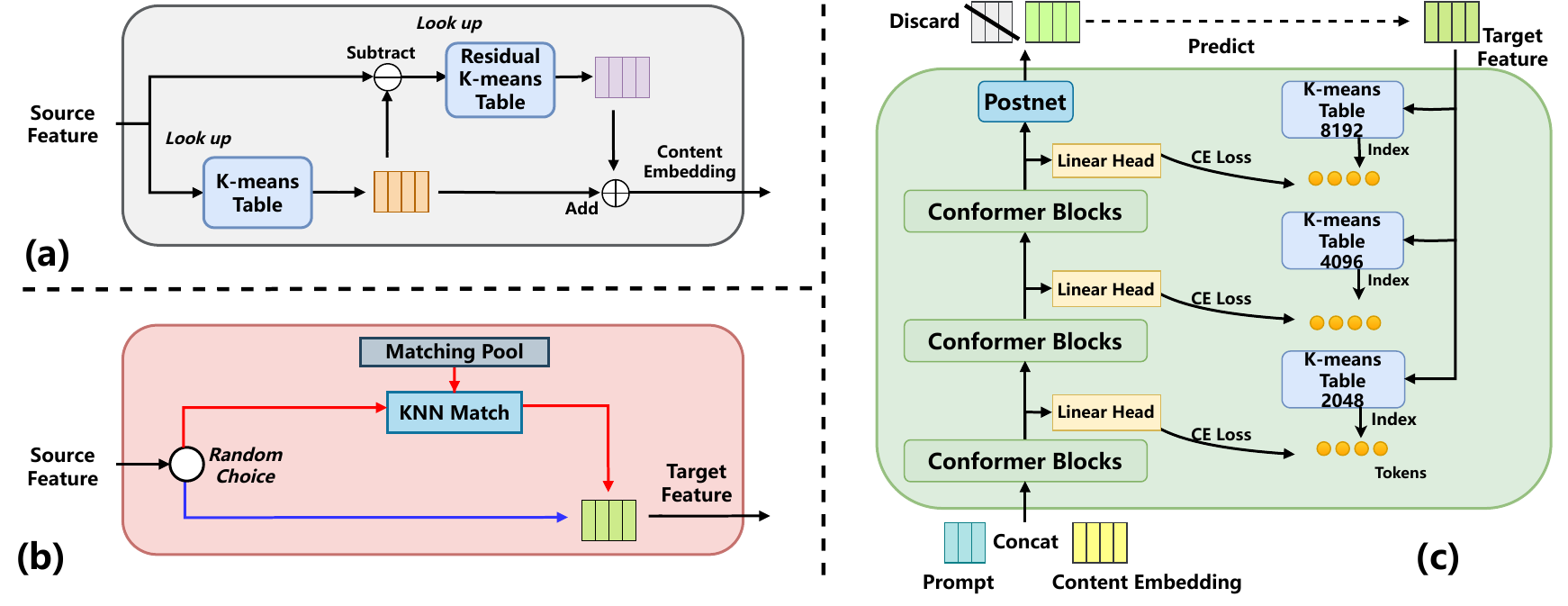}
    \caption{The details of Vec-Tok-VC+. (a): the residual-enhanced K-Means decoupler. (b): the dual-mode teacher guidance module. (c): the converter and multi-codebook progressive constraint.}
    \label{decouple_converter_teacher}
\end{figure*}

As mentioned in Sec.~\ref{sec:intro}, decomposing speech into different speech components is very essential to achieve zero-shot VC. Since the continuous SSL feature contains rich semantic and speaker timbre information, the common practice~\cite{vectok} is to set an information bottle via K-Means quantization to squeeze out the speaker timbre from the content information. 
But it usually hurts the linguistic information and causes the loss of speaking variations. This side effect makes the conversion results in potentially unnatural pronunciation and difficult to generate speech with rich speaking variation from source speech. To immigrate this problem, inspired by the mechanism of Residual vector quantization (RVQ), as shown in Fig.~\ref{decouple_converter_teacher} (a), rather than single K-Means clustering, we perform a residual-enhance clustering with two K-Means processes to make an enhanced content representation. Specifically, in the bottom of Fig.~\ref{decouple_converter_teacher} (a), the first 1024-category K-Means quantizes the source SSL feature to content representation. Using residual information between the raw continuous SSL feature and the quantize-after feature as input, the second 256-category K-Means in the residual path compensates the content representation with more linguistic information and speaking variation. This residual-enhanced content representation is used for further conversion. Notably, during this process, the quantize-after feature is represented by the centroid vectors, not the corresponding discrete indices.

\subsection{Dual-mode training with teacher-guided refinement}
\label{sec:teacher}
Most VC methods reconstruct speech during training, in which the speaker reference and source semantic content are both from the same utterance. But the speaker reference is provided by another utterance during conversion. This mismatch makes it hard to ensure the decoupling of speaker timbre and content information during training, causing potential performance degradation. To mitigate this problem, the intuitive way is to follow the conversion behavior but the ground truth of the converted speech does not exist in non-parallel datatset. Recently, kNN-VC has achieved remarkable few-shot any-to-any VC performance which replaces source SSL features with SSL features from the target speaker according to the feature similarity by k-nearest neighbors matching to achieve conversion. As shown in Fig.~\ref{decouple_converter_teacher} (b), inspired by kNN-VC, we introduce a teacher-guided refinement using a teacher module to simulate the conversion in training, called \textit{conversion mode}. To be specific, we collect speech utterances of 490 speakers each with approximately 7 minutes to form the matching pool. Each utterance is represented by an XLSR-based SSL feature. During the training of conversion mode, one target speaker is randomly selected from the matching pool, and the source feature is converted to form a pseudo-target feature with the speaker timbre by kNN matching. The converter is minimized by the generation loss between the predicted target feature and the pseudo target feature. In contrast, during the \textit{reconstruction mode}, the source feature and target feature are both from the same speech utterance. Notably, in this dual-mode training process, the conversion mode and reconstruction mode are randomly activated in 0.5. A 3-second slice is randomly selected from the output of the teacher module as the target speaker prompt.

\subsection{Prompt-based conformer converter}
With the decoupled content representation and target speaker utterance, the converter aims to capture the target speaker timbre and fuse it with source content to get the final conversion result. Following the recent advances~\cite{vectok}, the converter is achieved by a multi-layer conformer~\cite{conformer}, a variant of Transformer~\cite{transformer}, with prompt-based speaker modeling as presented in Fig.~\ref{decouple_converter_teacher} (c).
To be specific, the converter is designed as a non-autoregressive architecture composed of several conformer layers and a convolution-based postnet. 
Before inputting to the converter, the 3-second speaker prompt is concatenated ahead of the content embedding along the temporal axe. Benefiting from the in-context learning ability inherent in conformer, the converter can capture fine-grained speaker information and fuse it into conversion results with the source content. 
Mean square error (MSE) loss $\mathcal{L}_{mse}$ is used to measure the distance between the perdition feature and the target feature. A structural similarity loss~\cite{ssimloss} $\mathcal{L}_{ssim}$ is also introduced to ensure the generation quality. 
Furthermore, to better optimize the converter and prevent the information dispersed during the multi-layer modeling, we introduce a multi-codebook progressive constraint to help the model fit the target speech from coarse to fine details.

\textbf{Multi-codebook progressive constraint:} 
As can seen in Fig.~\ref{decouple_converter_teacher} (c), from the bottom layer to the top layer, 
the hidden output of the converter layer should also have an increased information richness~\cite{pvaetts} for transferring the content information to the complicated SSL features.  
To supervise this process and ensure content accuracy, we introduce a multi-codebook progressive constraint in hidden layers of the converter. Specifically, we perform three K-Means clustering on target features with small, medium, and large codebook numbers, respectively, which are 2048, 4096, and 8192 in practice. From the small to the large, the quantitation with different granularities can encode more diverse information about the speech. Thus, the quantize-after feature with a small codebook number is used to constrain the hidden output from the bottom layer and so on. Finally, this progressive loss $\mathcal{L}_{pro}$ is optimized by cross-entropy (CE) loss between the hidden outputs and the quantization results, which can be defined as $\mathcal{L}_{pro} = \mathcal{L}_{small} + \mathcal{L}_{meduim} + \mathcal{L}_{large}$.
The total loss functions of our system can be summarized as $ \mathcal{L}_{total} = \mathcal{L}_{mse} + \mathcal{L}_{ssim} + \mathcal{L}_{pro}$.

\section{Experiments}

\subsection{Experimental setup}

\textbf{Datasets: }
Our training set comprises a total of 19,000 hours of speech data, consisting of open-source English datasets LibriTTS~\cite{libritts} and Gigaspeech~\cite{gigaspeech}, and a Chinese audiobook dataset collected from internal resources. We preserve 80 English and 80 Chinese utterances as the test set.
We collect 10 English and 10 Chinese unseen speakers from outside the set as target speakers to evaluate the model's performance.

\noindent\textbf{Implement details:}
We utilize a pre-trained XLS-R\footnote{\hyperlink{https://pytorch.org/audio/stable/generated/torchaudio.models.wav2vec2_xlsr_300m}{https://pytorch.org/audio/stable/generated/torchaudio.models.wav2\\vec2\_xlsr\_300m}}model to extract 1024-dimension features with a 20 ms hop length from its sixth immediate layer as our speech representations.
All K-Means clustering is performed on the frame-level speech representations.
The k value for the average vector in the teacher module is 8.
The converter contains 6 Conformer blocks with 8 attention heads, an embedding dimension of 1024, a feed-forward layer dimension of 4096, and a dropout rate of 0.1, and a postnet consists of 4 layers of convolution with kernel size 5 and a dropout rate of 0.2.
The three prediction heads predict different granular semantic tokens of the target features from the output features of the 2nd, 4th, and 6th conformer block respectively.
A HiFiGAN V1 model is used as the vocoder, which takes speech representatives as the input and output 24kHz waveform. 
During training, we use Adam as the default optimizer with an initial learning rate of 0.0005, and $\beta_1$ = 0.9, $\beta_2$ = 0.95. 
Our conversion model is trained by 8 NVIDIA RTX3090 GPUs with a batch size of 8 per GPU for 500k steps.
The HiFiGAN model is trained by 4 NVIDIA RTX3090 GPUs with a batch size of 4 per GPU for 1,000k steps.

\noindent\textbf{Comparison models:}
We compare with two representative zero-shot VC systems: LM-VC~\cite{LMVC} and SEF-VC~\cite{SEF-VC}. 
The LM-VC adopts a two-stage framework with three LMs to achieve any-to-any VC.
The SEF-VC is a speaker-embedding-free voice conversion model that learns and incorporates speaker timbre from reference speech via a position-agnostic cross-attention mechanism and reconstructs waveforms from HuBERT semantic tokens non-autoregressively.
We train these two models on the same dataset as Vec-Tok-VC+ for fair comparison.

\noindent\textbf{Evaluation metrics:}
For subjective evaluation, the mean opinion score is used to measure speech naturalness (NMOS) and speaker similarity (SMOS) that are calculated with 95\% confidence intervals. Thirty participants with basic Chinese-English bilingual skills participated in the subjective experiments. Participants focus on specific aspects while disregarding others in the scoring process. 
Metrics for objective evaluation include speaker embedding cosine similarity (SECS), character error rate for Chinese (CER), and word error rate (WER) for English.
SECS is obtained by calculating cosine similarity between speaker embeddings extracted by Resemblyzer\footnote{\hyperlink{https://github.com/resemble-ai/Resemblyzer}{https://github.com/resemble-ai/Resemblyzer}}.
WER and CER for English and Chinese respectively are obtained by open-source ASR models based on the U2++ conformer architecture provided by the WeNet community~\cite{DBLP:conf/interspeech/YaoWWZYYPCXL21}.

\subsection{Zero-shot voice conversion results}

As shown in Table~\ref{vcResults}, Vec-Tok-VC+ achieves better naturalness and intelligibility than comparison models in intra-lingual zero-shot VC. We attribute this to the decoupling enhanced by the residual K-Means clustering and the constraint of the progressive loss function. 
Vec-Tok-VC+ achieves the best speaker similarity in both subjective and objective evaluations, indicating that the self-attention mechanism in our converter better captures and incorporates speaker information from 3-second XLS-R feature prompts. 
We conduct experiments on cross-lingual zero-shot VC to prove the capabilities of our model further. Although all models encounter a performance degradation during cross-lingual zero-shot VC, Vec-Tok-VC+ still outperforms comparison models. These results demonstrate the superiority of the Vec-Tok-VC+.

During the experiment, we find that Vec-Tok-VC+ also exhibits robust conversion capabilities for noisy source speech.
We show this ability through our demo page~\footnote{\hyperlink{https://ma-linhan.github.io/VecTokVC-Plus/}{https://ma-linhan.github.io/VecTokVC-Plus/}}.

\begin{table}[h]
\caption{Results of intra-lingual and cross-lingual zero-shot VC.}
\resizebox{\columnwidth}{!}{%
\begin{tabular}{cccccc}
\toprule[\heavyrulewidth]
Model   & NMOS$\uparrow$               & SMOS$\uparrow$               & CER$\downarrow$          & WER$\downarrow$          & SECS$\uparrow$           \\ \hline
GT      & 4.39$\pm$0.14          & -                  & 2.9          & 1.8          & -              \\ \hline
\multicolumn{6}{c}{\textit{intra-lingual vc}}                                                    \\
LM-VC   & 3.79$\pm$0.10          & 3.78$\pm$0.10          & 3.7          & \textbf{2.5} & 0.814          \\
SEF-VC  & 3.81$\pm$0.11          & 3.99$\pm$0.09          & 4.2          & 2.6          & 0.827          \\
Vec-Tok-VC+ & \textbf{3.98$\pm$0.11} & \textbf{4.05$\pm$0.12} & \textbf{3.4} & \textbf{2.5} & \textbf{0.841} \\ \hline
\multicolumn{6}{c}{\textit{cross-lingual vc}}                                                    \\
LM-VC   & 3.63$\pm$0.07         & 3.75$\pm$0.10          & 4.3          & 2.9          & 0.806          \\
SEF-VC  & 3.72$\pm$0.13          & 3.92$\pm$0.10         & 4.5          & 3.0          & 0.820          \\
Vec-Tok-VC+ & \textbf{3.90$\pm$0.08} & \textbf{3.99$\pm$0.08} & \textbf{3.5} & \textbf{2.6} & \textbf{0.836} \\ \hline
\end{tabular}%
}
\label{vcResults}
\end{table}

\subsection{Ablation study}
To compare with Vec-Tok-VC and investigate the importance of the methods we proposed, four ablation systems are obtained by dropping the teacher guidance module, the progressive loss function, the residual clustering in the decoupling process, and replacing the XLS-R with WavLM, respectively.
We denote them as \textit{-teacher module}, \textit{-progressive loss}, \textit{-residual cluster}, and \textit{*WavLM} respectively, as shown in Table~\ref{ablation}.
When dropping the teacher module, it only performs reconstruction during training and leads to an overall decline in performance.
The removal of progressive loss function brings performance decreases to both speech naturalness and speaker similarity.
Moreover, the removal of the residual clustering results in a significant decrease in naturalness, despite still maintaining a high level of speaker similarity, indicating a single-level K-Means captures insufficient linguistic content or prosodic details. 
Similarly, the replacement of XLS-R maintains the speaker similarity almost unchanged but brings a significant decrease in naturalness and intelligibility, showing the advantages of XLS-R in multi-lingual speech representation modeling.

\begin{table}[h]
\centering
\caption{Results of ablation study with 95\% confidence interval.}
\label{ablation}
\resizebox{\columnwidth}{!}{%
\begin{tabular}{llllll}
\toprule[\heavyrulewidth]
Model   & NMOS$\uparrow$ & SMOS$\uparrow$ & CER$\downarrow$ & WER$\downarrow$ & SECS$\uparrow$ \\ \midrule
Vec-Tok-VC+   &\textbf{3.93$\pm$0.09}      &4.01$\pm$0.10      &\textbf{3.5}     &\textbf{2.5}     &\textbf{0.839}      \\
\textit{-teacher module}  &3.83$\pm$0.06      &3.83$\pm$0.14      &3.6     &\textbf{2.5}     &0.808      \\
\textit{-progressive loss}  &3.80$\pm$0.11      &3.96$\pm$0.10      &4.1     &2.9     &0.823      \\
\textit{-residual cluster}  &3.71$\pm$0.11      &\textbf{4.02$\pm$0.12}      &4.9     &3.7     &\textbf{0.839}      \\
\textit{*WavLM} &3.72$\pm$0.08      &3.99$\pm$0.13      &3.9     &2.8     &0.831      \\ 
\bottomrule[\heavyrulewidth]
\end{tabular}%
}
\end{table}

\section{Conclusion}
In this paper, we proposed a prompt-based robust zero-shot VC model Vec-Tok-VC+. Improved from Vec-Tok Codec, it decouples content by residual-enhanced K-Means quantization on XLS-R representations and models speaker information given a 3-second speaker prompt by built-in self-attention. Besides, teacher-guided refinement achieved by kNN matching is introduced to simulate the conversion behavior to form a dual-mode training strategy to eliminate the training-inference mismatch for better conversion performance. Furthermore, a multi-codebook progressive loss function is used to constrain the layer-wise output of the model from coarse to fine during training.
Experiments and ablations demonstrate the superior performance and effectiveness of our proposed method.


\bibliographystyle{IEEEtran}
\bibliography{mybib}

\end{document}